\input{aipcheck}

\documentclass[
    ,final            
  ]
  {aipproc}

\layoutstyle{6x9}

\begin{document}

\title{Physics of {P}sychophysics: it is critical to sense}



\author{Mauro Copelli}{
  address={Laborat\'orio de F\'{\i}sica Te\'orica e Computacional \\
           Departamento de F\'{\i}sica, Universidade Federal de  Pernambuco\\
           50670-901, Recife, PE, Brazil}
}


\begin{abstract}
It has been known for about a century that psychophysical response
curves (perception of a given physical stimulus vs. stimulus
intensity) have a large dynamic range: many decades of stimulus
intensity can be appropriately discriminated before saturation. This
is in stark contrast with the response curves of sensory neurons,
whose dynamic range is small, usually covering only about one
decade. We claim that this paradox can be solved by means of a
collective phenomenon. By coupling excitable elements with small
dynamic range, the {\em collective\/} response function shows a much
larger dynamic range, due to the amplification mediated by excitable
waves. Moreover, the dynamic range is optimal at the phase transition
where self-sustained activity becomes stable, providing a clear
example of a biologically relevant quantity being optimized at
criticality. We present a pedagogical account of these ideas, which
are illustrated with a simple mean field model.
\end{abstract}

\maketitle


\section{The dynamic range problem}

\subsection{Introduction}

Physical stimuli impinge on our senses with a  range of
intensities that spans several orders of magnitude. 
How can animals cope with that scenario?  In order for them to
survive, their brains have to be able to distinguish among very weak
input signals, as well as among very strong ones.  This ability to
distinguish among widely varying signals can be formalized in
different ways, usually involving information theory as the main
conceptual tool.
Here we are going to employ an intuitive and much simpler concept to
embody this ability: the dynamic range.

\begin{figure}
  \includegraphics[width=.6\textwidth]{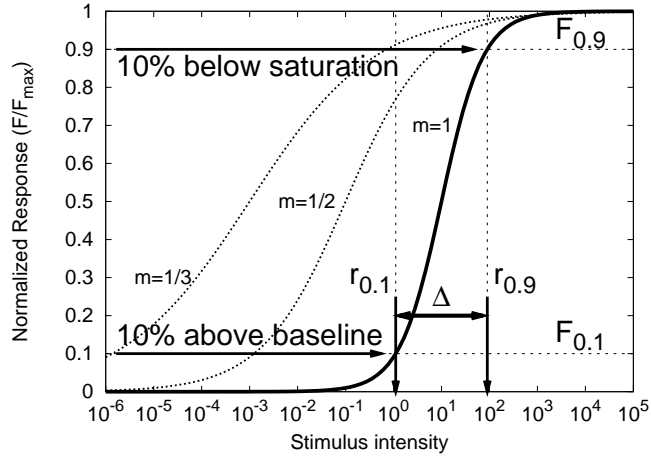}
  \caption{\label{fig:resposta}Linear saturating curve, corresponding
  to (normalized) Hill functions (Eq.~\ref{eq:Hill}) with exponents
  $m=1$ (solid), $m=1/2$ and $m=1/3$ (dashed).}
\end{figure}

Consider for instance, the response curve labeled $m=1$ in
Fig.~\ref{fig:resposta}. The horizontal axis (in log scale) represents
the stimulus intensity $r$ (for instance, the concentration of an
odorant inside your nose, or the intensity of light reaching your
retina), whereas the vertical axis is the response $F$ to that
stimulus. Suppose $F$ represents the mean firing rate of some early
sensory layers of the nervous system, which are responsible for the
initial transduction from a physical stimulus to neural activity. This
neural activity will presumably be ``read'' by other neurons in higher
areas of the brain, which will further process it, and so on. What
those higher areas ``see'' is therefore $F$, from which one could in
principle infer $r$ by taking the inverse of the response function
$F(r)$. Note, however, that for the $m=1$ curve in
Fig.~\ref{fig:resposta}, it would be difficult to perform such an
inversion operation if the stimulus was very weak, say, $r\sim
10^{-5}-10^{-4}$. The reason is that $F$ is very close to a plateau at
its baseline activity $F_0 \equiv \lim_{r\to 0}F(r)$ ($=0$, in this
example). The same difficulty would arise for very strong stimulus
(say, $r\sim 10^{4}-10^{5}$ in Fig.~\ref{fig:resposta}), in which case
$F$ is very close to its saturation plateau $F_{max}=\lim_{r\to\infty}
F(r)$.

To exclude these regions, the dynamic range $\Delta$ (measured in dB)
is defined as $\Delta =
10\log_{10}\left(r_{0.9}/r_{0.1}\right)$,
where the range $[r_{0.1},r_{0.9}]$ is obtained from the response
interval $[F_{0.1},F_{0.9}]$, as illustrated in
Fig.~\ref{fig:resposta}. To estimate the range of stimuli that can be
discriminated, one simply discards stimuli which are too faint to be
detected ($r<r_{0.1}$) or too close to saturation
($r>r_{0.9}$).
This is clearly an arbitrary choice, but it is usual in the biological
literature and very useful as an operational definition. To account
for systems which have a nonzero baseline activity $F_0$, the general
definition of $F_{x}$ is simply~\cite{Kinouchi06a}


\begin{equation}
\label{eq:Fx}
F_x = F_0 + x(F_{max}-F_{0})\; .
\end{equation}
In the case of the $m=1$ curve in Fig.~\ref{fig:resposta}, the dynamic
range is about 19~dB, i.e. almost two decades. Therefore, if a system
had such a response curve, it would have a hard time handling stimulus
intensities varying by more than two decades (as natural stimuli do).


\subsection{Psychophysics: large dynamic range}


The fact that animals {\em can\/} operate with a wide dynamic range is
most easily revealed in humans by classical results in
Psychophysics~\cite{Stevens}: the {\em perception\/} of the intensity
of a given stimulus is experimentally shown to depend on the stimulus
intensity $r$ as $\sim \log(r)$ (Weber-Fechner law) or $\sim r^m$
(Stevens law), where the Stevens exponent $m$ is usually~$<1$. Those
empirical laws are known for about a century and have in common the
fact that their dynamic range is large. You can convince yourself that
small exponents lead to large dynamic ranges by looking again at
Fig.~\ref{fig:resposta}, which shows Hill functions with different
exponents $m$:
\begin{equation}
\label{eq:Hill}
F(r) = \frac{F_{max} r^m}{r_0^m + r^m}\; .
\end{equation}
Notice that the Hill function can be thought of as a saturating
Stevens law, both sharing the same exponent $m$ for low stimulus. It
is a simple exercise to show that the dynamic range dependence on $m$
for the Hill function is $\Delta = \frac{10}{m}\log_{10}(81)$.

\subsection{Sensory neurons: small dynamic range}

Let us focus on one particular sense, namely, olfaction.  The dynamic
range problem becomes evident when one looks at the experimental
response curve of olfactory sensory neurons (OSNs), which are at the
very first stage of signal processing and translate odorant
concentration into firing rates: their dynamic range is {\em small\/},
typically about 1 or 2 decades only~\cite{Rospars00}! That is: on the
one hand, sensory neuron responses look like the $m=1$ curve in
Fig.~\ref{fig:resposta}; on the other hand, psychophysical responses
look like the $m<1$ curves in Fig.~\ref{fig:resposta}. How can we
reconcile those results? How could psychophysical laws with large
dynamic range be physically implemented if individual sensory neurons
have a small dynamic range? How do exponents $<1$ arise
from the dynamics of neurons?

\begin{figure}
  \includegraphics[width=.6\textwidth]{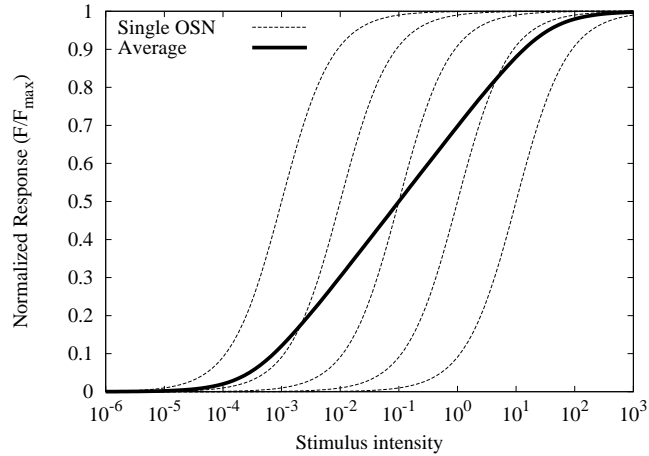}
  \caption{\label{fig:cleland}A simplified picture of recruitment
  theory. Each dashed curve is a Hill function with $m=1$ but
  different sensitivity $r_0$ (see Eq.~\ref{eq:Hill}). The average of
  the five curves has a much larger dynamic range.}
\end{figure}

One theory that tries to explain such a discrepancy has been proposed
by Cleland and Linster~\cite{Cleland99}. Their idea lies on the
presumed heterogeneity within the population of OSNs with the same
type of odorant receptor. If some OSNs have more (less) receptors on
their surface, they'll be more (less) sensitive and their response
curve will saturate earlier (later), like the dashed curves on the
leftmost (rightmost) part of Fig.~\ref{fig:cleland}. As the odorant
concentration gradually increases, more and more neurons would be
``recruited'', so that the average response (solid curve) would have a
large dynamic range even if each of the neurons had a small dynamic
range. Appealing at first sight, the problem with such a ``recruitment
theory'' is that for each order of magnitude in sensitivity, one needs
an order of magnitude in receptor density. Experimentally, however,
receptor over-expression is only about twofold~\cite{Cleland99}, so it
is plausible to assume that this is not the main mechanism responsible
for the phenomenon.


\section{A Collective Solution}

In recent years we have been working on a different solution to the
dynamic range
problem~\cite{Copelli02,Copelli05b,Copelli05a,Furtado06,Kinouchi06a}. The
idea is that by coupling excitable elements with small dynamic range,
one obtains an excitable medium whose response function will have a
large dynamic range due to a collective phenomenon. In order to build
a simple model of this mechanism, let us first study a toy model of a
single excitable element.

\subsection{Response function of a single excitable element}

Olfactory sensory neurons behave as excitable elements. In the absence
of an external stimulus, they essentially stay quiet. They will spike
if odorant molecules with enough affinity bind to the receptors on
their surface. After spiking, they undergo a refractory period before
they can spike again. The stronger the odorant concentration, the more
likely (on average) this process will repeat itself.

Consider a simple Greenberg-Hastings cellular automaton model, where
each excitable element $i=1,\ldots,N$ has $n$ states: $s_i=0$ is a
resting state (polarized neuron), $s_i=1$ is an excited state (spiking
neuron) and $s_i=2,\ldots,n-1$ are refractory states (hyperpolarized
neuron). The rules are as follows: $s_i=0$ changes to $s_i=1$ in the
next time step only if a supra-threshold stimulus arrives, otherwise it
does not change. If $s_i(t)\geq 1$, then $s_i(t+1) = [s_i(t) +
1]\mbox{mod}\; n$, that is: after an excitation, the element goes
through $n-2$ refractory states (blind to new stimuli) before
returning to $s_i=0$.
We model the arrival of stimuli by a Poisson
process: the probability for an element to jump from $s_i=0$ to $s=1$
is $\lambda(r) = 1-\exp(-r\delta t)$, where $r$ is assumed to be
proportional to the odorant concentration and the time step $\delta
t=1$~ms sets the time scale of the model.


Notice that we have an ensemble of excitable elements which are not
coupled to one another, so the problem can be solved exactly. If we
denote by $P_t(k)$ the probability that we find an element in state
$k$ at time $t$, then the rules stated above immediately yield:
\begin{eqnarray}
\label{eqs}
P_{t+1}(1) & = & \lambda P_t(0) \nonumber \\
P_{t+1}(2) & = & P_t(1) \nonumber \\
 & \vdots & \nonumber \\
P_{t+1}(n-1) & = & P_t(n-2) \; .
\end{eqnarray}
To obtain $P_t(0)$ we make use of the normalization condition
$\sum_{k=0}^{n-1}P_t(k) = 1$. To obtain the response function in the
stationary state, we take the limit $t\to\infty$. Dropping the $t$
index in the probabilities, Eqs.~\ref{eqs} lead to $P(n-1)= P(n-2) =
\ldots = P(1) = \lambda P(0)$. Normalization then leads to
\begin{equation}
\label{norm}
P(0)=1-(n-1)P(1)\; . 
\end{equation}
Solving for $P(1)$, we obtain the response
function~\cite{Copelli02,Furtado06}:

\begin{equation}
\label{desacoplado}
P(1) = F(r) \delta t = \frac{\lambda(r)}{1+(n-1)\lambda(r)}\; .
\end{equation}
We omit $\delta t=1$~ms from now on.  Since $\lambda(r)$ is
approximately linear for small $r$, Eq.~\ref{desacoplado} is similar
(but not identical) to a Hill function with $m=1$. The reader is
invited to show that the dynamic range is $\Delta =
10\log_{10}\{\ln[1+9/n]/\ln[1+1/(9n)] \}$, which is a smooth function
of $n$ that quickly saturates at $10\log_{10}(81)\simeq 19$~dB. This
very simple model correctly reproduces the experimental fact that
isolated OSNs have small dynamic range.

\subsection{Coupled excitable elements: mean field results}

All OSNs expressing the same receptor send their axons to the same
glomerulus, where axon terminals meet the dendritic trees of about
twenty mitral cells. Those dendrites are believed to be active, so
each dendritic patch can be modeled as an excitable element. Moreover,
it has recently been demonstrated that gap junctions (electrical
synapses) exist among mitral cell dendrites~\cite{Kosaka05}. What we
would like to show is that the interaction among those excitable
elements collectively lead to an enhancement of sensitivity {\em
and\/} dynamic range.

\begin{figure}
  \includegraphics[width=.51\textwidth]{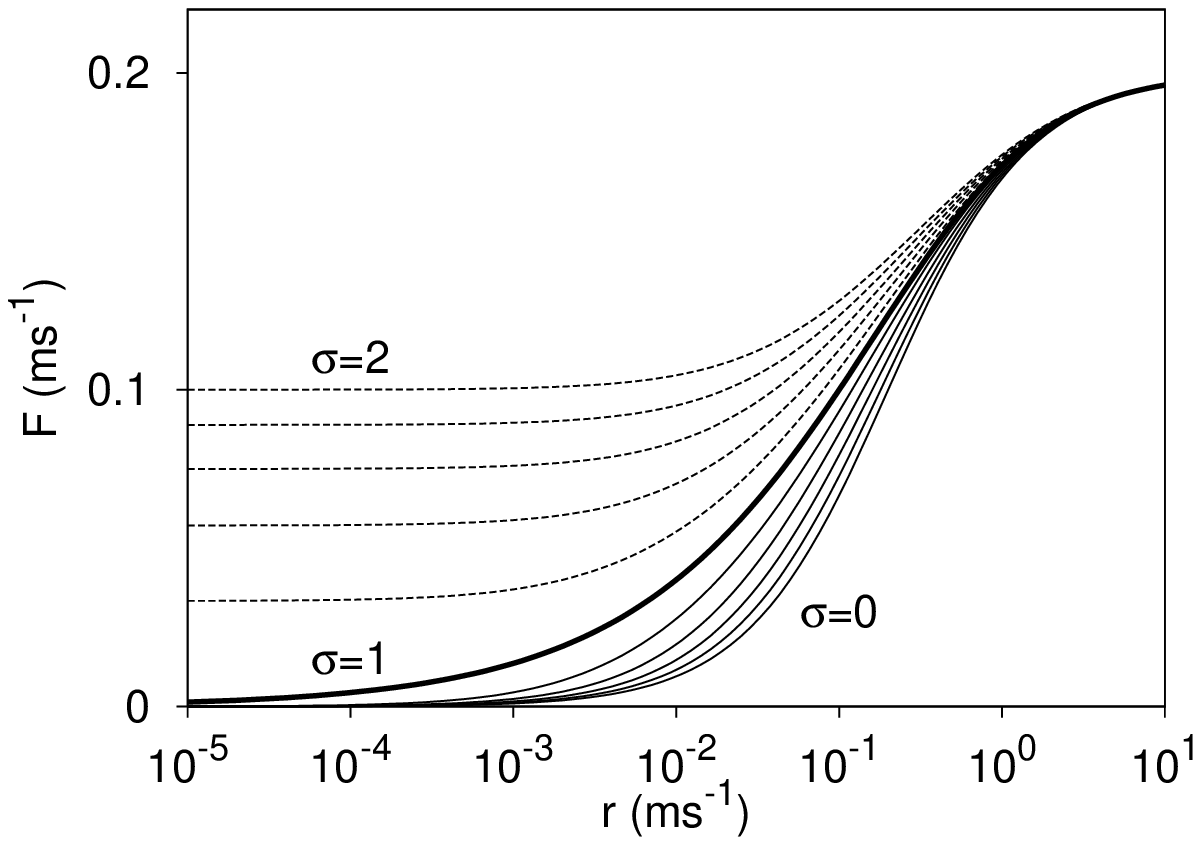}
  \includegraphics[width=.51\textwidth]{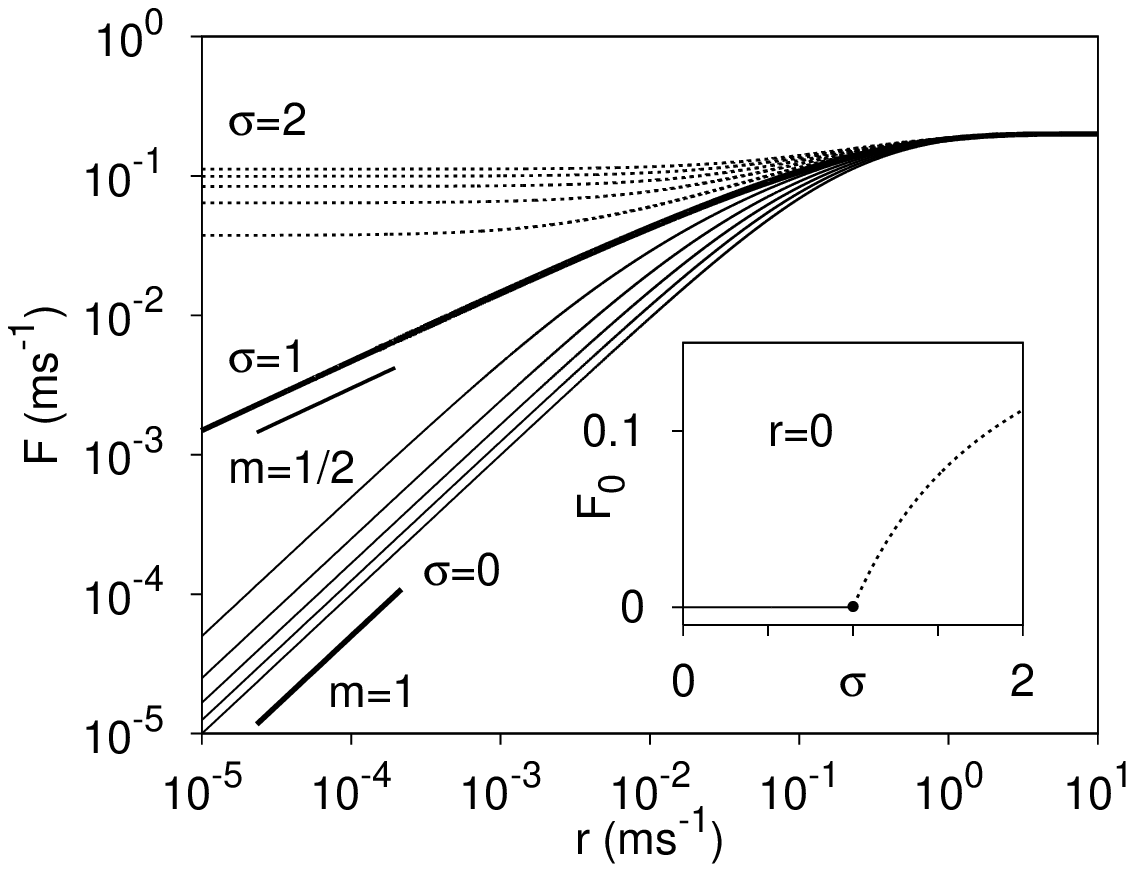}
  \caption{\label{fig:responses}Subcritical (thin solid), critical
  (thick solid) and supercritical (dashed) response curves for the
  mean field model in linear-log (left) and log-log (right)
  scales. Inset: self-sustained activity without external stimulus.}
\end{figure}

Let us study a very simple model~\cite{Kinouchi06a}.
First, we assume that, owing to the gap junctions, a dendritic patch
is randomly coupled to $K$ other patches, each one modeled by our
simple cellular automaton and independently subjected to external
stimuli with probability $\lambda(r)$. Furthermore, the coupling is
such that an excitation at one site can propagate with probability $p$
to its quiescent neighbors. In a mean field description, the
stationary probability that a site is in the excited state is $P(1) =
F = P(0)\left[1 - (1-\lambda)(1-pF)^K \right]$,
where the last parenthesis is the probability that no excitation comes
from the $K$ neighbors (a fraction $F$ of which are likely to be
active). Together with the normalization condition in Eq.~\ref{norm},
one arrives at the self-consistent equation for the response function

\begin{equation}
\label{respostaMF}
F(r) = \left(1-(n-1)F\right)\left[1 - (1-\lambda(r))(1-pF)^K \right]\; .
\end{equation}

\begin{figure}
  \includegraphics[width=.6\textwidth]{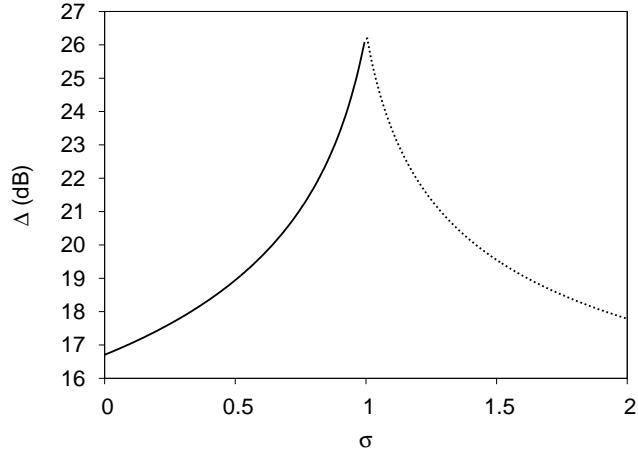}
  \caption{\label{fig:deltasigma}Optimal dynamic range is obtained at
  the critical value $\sigma_c=1$.}
\end{figure}

The solution of Eq.~\ref{respostaMF} for $K=10$ and $n=5$ is plotted
in Fig.~\ref{fig:responses}. Our control parameter is the branching
ratio $\sigma\equiv pK$, which is approximately the average number of
excitations transmitted by an excited site to its neighbors. Starting
from $\sigma=0$, we see that increasing $\sigma$ leads to amplified
response of low stimuli due to propagation of excitable waves, so the
dynamic range {\em increases\/} (Fig~\ref{fig:deltasigma}). Then, at
$\sigma=\sigma_c \equiv 1$ a nonequilibrim phase transition occurs!
For $\sigma > 1$, each site is transmitting more excitations than it
is receiving, so it's not surprising that even in the absence of
external stimulus ($r\to 0$) the system is able to maintain a
self-sustained activity (see inset of Fig.~\ref{fig:responses}). If we
keep increasing $\sigma$ above criticality, this self-sustained
activity masks the response to weak stimuli, so the dynamic range {\em
decreases} (recall the effect of $F_0$ in Eq.~\ref{eq:Fx}). Therefore,
{\em the maximum dynamic range is obtained precisely at
criticality\/}.

Another signature of criticality is the power law behavior of the
response curve (Fig.~\ref{fig:responses}, right). For $\sigma<1$, the
weak stimulus response is linear, $F \sim r$. But for
$\sigma=\sigma_c$, the response is $F \sim r^{1/2}$, as can be easily
verified by expanding Eq.~\ref{respostaMF} around $F=0$. What is
remarkable is that this exponent $1/2$ at criticality is very close to
the measured Stevens exponents for light and odor intensity ($m=0.5$
and $0.6$, respectively~\cite{Stevens}). We claim that Stevens law is
a power law because our sensory systems should be critical. The
motivation for being critical is clear: it allows the system to
operate with high sensitivity and large dynamic range, both of which
are desirable features for a brain living in a world ``where extreme
events exist, and where probabilities often have long
tails''~\cite{Chialvo06}.

Experimentally, glomeruli have larger dynamic ranges than
OSNs~\cite{Wachowiak01} (which was in fact what motivated the
model). The hypothesis that the propagation of activity in the
glomerulus is dominated by gap junctions could be tested by measuring
the response curve in mice where Connexin-36 (the protein that
accounts for the gap junctions) has been genetically knocked out (in
fact, analogous experiments in the retina are consistent with this
idea~\cite{Furtado06}). But clearly the mechanism we propose is not
exclusive of electrically coupled systems, being a rather general
property of excitable media.

\section{Concluding remarks}

Those familiar with nonequilibrium phase transitions will recognize
the response exponent $m$ at criticality as the critical exponent
often named $1/\delta_h$ (see~\cite{Marro99}), and $1/\delta_h=1/2$ is
just the well known mean field value. While the simple model we have
presented seems well suited to describe an apparently disordered and
highly interwoven structure like the olfactory
glomerulus~\cite{Kinouchi06a}, one can go beyond mean
field~\cite{Furtado06}. In excitable media with a different topology
it is fair to expect that exponents will belong to the Directed
Percolation (DP) universality class (even though this is not always
the case~\cite{Furtado06}). If one looks at DP in hypercubic networks,
for instance, $1/\delta_h$ is always $\leq 1/2$. In this sense, the
mean field results for optimal dynamic range can be regarded as a {\em
lower bound\/}. Networks with a different structure could easily
surpass the peak at $26$~dB of Fig.~\ref{fig:deltasigma}.

To summarize, we have presented a framework where psychophysical laws
with large dynamic range emerge naturally from the interactions among
excitable elements with small dynamic range. In particular, both the
dynamic range and the sensitivity are optimized if the system is at
the phase transition where self-sustained activity becomes stable. We
point out that the dynamic range is an interesting observable, since
it is dimensionless, easy to measure and of great biological
relevance. The fact that it is maximized at a phase transition
provides a clear example of optimal information processing at
criticality, therefore building on a long history of efforts (both
theoretical and experimental) along the same idea~\cite{Langton90,
Bak, Beggs03, Chialvo04}.



\begin{theacknowledgments}
Mauro Copelli acknowledges financial support from FACEPE, CNPq and
special program PRONEX, as well as stimulating discussions with Dante
R. Chialvo.
\end{theacknowledgments}



\bibliographystyle{aipproc}   

\bibliography{copelli}

\IfFileExists{\jobname.bbl}{}
 {\typeout{}
  \typeout{******************************************}
  \typeout{** Please run "bibtex \jobname" to obtain}
  \typeout{** the bibliography and then re-run LaTeX}
  \typeout{** twice to fix the references!}
  \typeout{******************************************}
  \typeout{}
 }

\end{document}